\documentclass[a4paper,12pt]{article}
\usepackage[latin1]{inputenc}
\usepackage{natbib}
\usepackage{amssymb}
\usepackage{amsmath}
\usepackage{amsthm}
\usepackage{latexsym}
\usepackage{graphicx}
\usepackage{bm}
\usepackage{overpic}
\usepackage[normalem]{ulem}

\usepackage{exscale}
\usepackage{amsfonts}
\usepackage[usenames,dvipsnames]{color} 

\def\Om{\Omega}
 \def\ua{\uparrow}
 \def\da{\downarrow}

 \def\wh{\widehat}
 \def\wt{\widetilde}

\def\ignore#1{}

\def\bbar{\overline}
\def\bR{\mathbb{R}}

\def\cC{\mathcal C}

\def\cX{\mathcal X}
\def\cF{\mathcal F}
\def\bP{\mathbb P}
\def\bE{\mathbb E}
\def\bN{\mathbb N}

\def\eps{\varepsilon}

\def\cXsem{\cX_{\text{\rm sem}}}
\def\cXBV{\cX_{\text{\rm BV}}}

\def\Proof{\bigskip\noindent{\bf Proof. }}

\newtheorem{theorem}{Theorem}
\newtheorem{lemma}{Lemma}
\newtheorem{corollary}{Corollary}

\theoremstyle{definition}
\newtheorem{definition}{Definition}

\newtheorem{remark}{Remark}

\def\eins{{\mathchoice {1\mskip-4mu\mathrm l}
{1\mskip-4mu\mathrm l}{1\mskip-4.5mu\mathrm l}
{1\mskip-5mu\mathrm l}}}

\renewcommand{\baselinestretch}{1}\normalsize

\begin{document}
\title{\LARGE\bf  Drift dependence of optimal  trade  execution strategies under transient price impact}
\author{\normalsize Christopher Lorenz, Alexander
Schied\thanks{Support by Deutsche Forschungsgemeinschaft is gratefully acknowledged.}\\ \normalsize Department of Mathematics\\ \normalsize University of Mannheim\\
\normalsize A5, 6, 68131 Mannheim, Germany\\
\normalsize{\tt chlorenz@uni-mannheim.de, schied@uni-mannheim.de}
}

\date{\small First version: January 26, 2012\\ This version: March 3, 2013}

\maketitle

\abstract{We give a complete solution to  the problem of minimizing the expected liquidity costs in presence of a general drift when the underlying   market impact model has linear transient price impact with exponential resilience. It turns out that this problem is well-posed only if the drift is absolutely continuous. Optimal strategies often do not exist, and when they do, they depend strongly on the derivative of the drift. Our approach uses elements from singular stochastic control, even though the problem is essentially non-Markovian due to the transience of price impact and the lack in Markovian structure of  the underlying price process. As a corollary, we  give a complete solution to the minimization of a certain cost-risk criterion in our setting. }

\section{Introduction}

Standard asset pricing models like the Black--Scholes model assume that asset prices are given exogenously and are unaffected by the trading behavior of economic agents. In reality, however, many trades are large enough to feed back on asset prices so that  {price impact} and the resulting liquidity costs cannot be ignored.
In such a situation, one aims at minimizing the liquidity costs from  trade  execution by constructing suitable trading strategies. The problem of computing such trading strategies is called the \emph{optimal  trade  execution problem}.

To deal with price impact quantitatively, several stochastic \emph{market impact models} have been proposed in recent years. In the first model class, which goes back to \cite{BertsimasLo} and \cite{AlmgrenChriss1,AlmgrenChriss2}, price impact is modeled by combining convex transaction costs with a linear permanent price impact term.  While these models make computations feasible and lead to relatively nice and robust trading strategies, they do not adequately model the empirically observed transience of price impact. Transience means that price impact is strongest immediately after being triggered  and that it subsequently decays in time. This effect is well-established empirically, it can be measured, and it is widely believed that the decay of price impact follows some general laws; see, e.g., \cite{Gatheral}, \cite{LehalleDang}, \cite{Moroetal}, and the references therein.
Therefore, several models for transient price impact have been proposed in recent years. To our knowledge, the first models were proposed by \cite{Bouchaudetal} and \cite{ow}. The latter is a linear price impact model with exponential decay of price impact and seems to be the first transient-price impact model  used for computing optimal  trade  execution strategies. Two different extensions were given to the case of nonlinear transient price impact. The first was proposed by \cite{AFS2} and further developed by \cite{AS} and \cite{PredoiuShaikhetShreve}. The second extension is due to \cite{Gatheral} and, besides nonlinearity, also allows for more general decay patterns than exponential decay. Let us also mention related research by \cite{BayraktarLudkovski}, \cite{BouchardDangLehalle},  \cite{KharroubiPham}, and \cite{GueantLehalleTaipa}.

Since transience of price impact is more realistic than the combination of transaction costs with linear permanent impact, one might guess that market impact models with transient price impact perform better in practice than those of \cite{BertsimasLo} and \cite{AlmgrenChriss1,AlmgrenChriss2}. But what can be said about their mathematical stability and robustness in comparison to these older models? This is an important question because of the high degree of uncertainty in the estimation of market microstructure parameters.  \cite{Gatheral} addressed this question by analyzing the  possible non-existence of optimal  trade  execution strategies for certain parameters. As shown by \cite{AS} and further discussed in \cite{GSS2}, these results depend strongly on the way in which nonlinearity of price impact is modeled. Therefore stability investigations with respect to other model features have been carried out in the case of linear price impact. Moreover, for liquid stocks linear price impact can also be a very good approximation to reality as shown empirically by \cite{BlaisProtter}. \cite{ASS} investigate the dependence of optimal  trade  execution strategies on the decay kernel that models the temporal  decay of price impact.  They find that discrete-time strategies react in a very sensitive manner to the choice of this decay kernel and that price impact must decay as a convex nonincreasing function of time so as to exclude certain irregularities of optimal strategies. This observation implies in particular that in practice the decay of price impact cannot be estimated in a nonparametric way.

An extension  of the results in \cite{ASS} to  continuous time  was given by \cite{GSS}. Finally, assuming exponential decay of price impact, \cite{FruthSchoenebornUrusov} analyze the specific form and regularity of optimal  trade  execution strategies when liquidity can be time-dependent or even stochastic. An analysis pertaining specifically to regularity issues arising in this context has recently been given by \cite{Kloeck}.

When investigating a particular model aspect, it is important   to keep the remaining features of the model  simple.  For instance, to analyze the existence or nonexistence of price manipulation strategies as in \cite{Gatheral} or \cite{ASS},  it is necessary to assume that the underlying price process is a martingale.  There are additional reasons why it may be natural to make this martingale assumption; see, e.g., the discussion in \cite{ASS}. But there are also good reasons to allow for a nonvanishing drift in unaffected asset prices.  For instance, an economic agent may be aware of the trading activities of another market participant.  These trading activities will create price impact, which from the point of view of our economic agent will be perceived as a drift in asset prices. Moreover, for several reasons, the economic agent may have a rather accurate estimate of this drift. For instance, some trade execution algorithms create characteristic order patterns and therefore allow for an inference of  their future trading trajectory. We refer to \cite{SchoenebornSchied}  for a study of a multi-agent situation in the Almgren--Chriss framework.

In this paper, we aim at continuing the investigation of the stability of  models for transient price impact by focusing on the dependence of optimal  trade  execution strategies on a possible \emph{drift} of the underlying unaffected price process. In doing this, we will allow for rather general dynamics of the drift and in particular allow for jumps and a non-Markovian structure. This is important because the price impact patterns of optimal  trade  execution strategies with transient price impact have precisely these features and, as mentioned above, the price impact of another market participant is perhaps the most common source for the presence of a drift. On the other hand, we will keep the remaining features of the model  simple. This makes  the mathematics  tractable  but also helps to isolate the effects of the drift from the effects created by other model features.
We therefore use the linear continuous-time  model of \cite{ow} (in the version of \cite{GSS}) with exponential decay of price impact and the problem we are looking at is the minimization of the expected costs.

Theorem~\ref{GenSemThm}, our main result, shows that this optimal  trade  execution problem  is very sensitive
with respect to the drift. The expected costs will be equal to negative infinity as soon as the drift is not absolutely continuous, a fact that will have strong impact when market impact is generated by several market participants.  Moreover, even when the drift is absolutely continuous, optimal strategies will typically not exist if strategies are understood in the sense of \cite{GSS}. We therefore extend the class of admissible strategies by allowing strategies to be semimartingales. We show that unique optimal  trade  execution strategies may exist in this class of strategies, but the number of shares to be held depends directly on the derivative of the drift at each time and thus may fluctuate strongly.
This sensitivity of strategies is particularly striking when compared to the relatively robust drift dependence of optimal  trade  execution strategies in the Almgren--Chriss framework, which was found by \cite{SchiedRobust}.

Our problem of minimizing the expected costs in the presence of a drift turns out to be also of interest from a purely mathematical point of view. Our approach uses elements from singular stochastic control, although the problem is basically non-Markovian due to both the transience of price impact and the lack in Markovian structure of  the underlying price process.  We deal with the first type of non-Markovianity by using an auxiliary \lq impact process\rq\ $E^X_t$ that, under the specific assumption of exponential decay of price impact,
leads to a Markovian structure for the dynamics of transient price impact.  We then guess a formula for the optimal expected costs conditional at time $t\ge0$ where
an arbitrary impact $E^X_t$ is given as initial condition.  With this formula {at hand}, we can then use a verification argument. The control problem is \lq singular\rq\ since our controls are semimartingale strategies, which enter the value function as integrators of stochastic integrals. A similar technique was recently used in \cite{AS12} to compute optimal strategies for general, completely monotone decay kernels but without drift in the unaffected price process. As an application of our results, we  also obtain a complete solution  for the minimization of a cost-risk criterion that was recently proposed in \cite{GatheralSchied}.

\section{Statement of results}

\subsection{Model setup}

A market impact model is a model for an economic agent who can move asset prices. As long as this agent is not active, asset prices are  determined by the actions of the other market participants and
are described by the \emph{unaffected price process} $S^0$. We assume that $S^0$ is a square-integrable c\`adl\`ag semimartingale defined on a given filtered probability space $(\Om,\cF,(\cF_t),\bP)$ {satisfying the usual conditions. We also assume that $\cF_0$ is $\bP$-trivial, i.e., every $\cF_0$-measurable random variable is $\bP$-a.s. constant.} We will use the linear market impact model with exponential decay of price impact  proposed by \cite{ow}. More precisely, we will use the zero-spread version of this model that was suggested in \cite{GSS}; we refer to \cite{AS} for a discussion of the possible re-introduction of a bid-ask spread.

The actual asset price will depend on the strategy chosen by the  trader. Such a strategy will be an adapted stochastic process $X=(X_t)_{t\ge0-}$ that describes the number of shares held by the trader at each time. Following  \cite{GSS}, we call $X$  \emph{admissible} if the following conditions are satisfied:
\begin{enumerate}
\item the function $t\rightarrow X_t$  is right-continuous\footnote{Although \cite{GSS} consider the left-continuous modification of $X$,  our definitions of both price process and costs coincide with the one in \cite{GSS}. See Remark~\ref{right-leftcont remark} for a detailed discussion of right versus left continuity.} and  adapted;
\item the function $t\rightarrow X_t$ has   finite and $\bP$-a.s. bounded total variation;
\item there exists  a \emph{liquidation time} $T\ge0$ such that $X_t=0$ $\bP$-a.s. for all $t\ge T$.
\end{enumerate}
Such a strategy has the interpretation that the value $X_{0-}$ stands for an initially given amount of shares that needs to be liquidated by time $T$. When $X$ is nonincreasing, it is a pure sell strategy. When it is nondecreasing, it is a pure buy strategy. A general admissible strategy is the sum of a sell and a buy strategy and therefore is of bounded variation. This shows that condition (b) is economically meaningful. With $\cXBV(x,T)$ we will denote the class of all strategies that are admissible in this sense for a fixed liquidation time $T\ge0$ and that satisfy $X_{0-}=x$.

When the admissible strategy $X$ is used, the  price $S^X_t$  will be
\begin{equation}\label{ModelEquation}
S^X_t=S^0_t+ \eta\int_{[0,t)}e^{-\rho(t-s)}\,dX_s,
\end{equation}
where  $\rho>0$, the function $e^{-\rho t}$ describes the temporal decay of price impact, and the parameter $\eta$ describes its magnitude. Clearly we can set $\eta:=1$ without loss of  generality.  Following \cite{GSS}, we define the liquidation costs of $X\in\cXBV(x,T)$ as
\begin{equation}\label{BV costs}
\cC(X):=\int_{[0,T]} S^0_t\,dX_t+\int_{[0,T]} \int_{[0,t)}e^{-\rho(t-s)}\,dX_s\,dX_t+\frac{1}2\sum_{t\in[0,T]}(\Delta X_t)^2.
\end{equation}

\begin{remark}[Economic motivation of the cost functional $\bm{\cC(\cdot)}$]\label{cost remark}
Let us follow \cite{ASS} and \cite{GSS} in motivating the cost functional \eqref{BV costs}.  For a continuous strategy $X\in \cX_{BV}(x,T)$, $\cC(X)$  equals $\int_0^TS^X_t\,dX_t$ and can thus be easily understood as the accumulated costs of buying $dX_t$ shares at price $S^X_t$ at each time $t$. For general $X$, a nonzero jump $\Delta X_t$ can be interpreted as a large market order which shifts the asset price by eating into a block-shaped limit order book. Its execution therefore incurs the following costs:
$$\int_{S^X_{t}}^{S^X_t+\Delta X_t}y\,dy=S^X_t\Delta X_t+\frac12(\Delta X_t)^2= S^0_t\Delta X_t+\int_{[0,t)}e^{-\rho(t-s)}\,dX_s\,\Delta X_t+\frac12(\Delta X_t)^2.
$$
We assume here that the order $\Delta X_t$ is executed immediately \emph{after} a jump of $S^0_t$ in case both jumps nominally occur at the same time, an assumption that is economically natural  since it precludes  arbitrage-like exploitation of price jumps. Decomposing a general strategy into its continuous part and its jumps thus leads to the definition \eqref{BV costs}. An alternative derivation of \eqref{BV costs}, based on a continuous-time limit of discrete-time cost functionals,  will be provided by  Lemma~\ref{SemimartingaleCostLemma} in the more general framework of semimartingale strategies.   \hfill$\diamondsuit$
\end{remark}

\medskip

The problem of minimizing the expected costs, $\bE[\,\cC(X)\,]$, over $X\in\cXBV(x,T)$ is called the \emph{optimal  trade  execution problem.}  When $S^0$ is a square-integrable martingale, this problem  admits the unique solution
\begin{equation}\label{ow solution}
X_t=\frac{x(1+\rho(T-t))}{2+\rho T},\qquad 0\le t<T.
\end{equation}
That is, $X$ has an initial jump at $t=0$ of size $\Delta X_0=\frac{-x}{2+\rho T}$, continuous trading at rate $dX_t=\frac{-x\rho}{2+\rho T}\,dt$ in $(0,T)$, and a terminal jump of size $\Delta X_T=\Delta X_0$. This formula was found by \cite{ow} (see also Example 2.12 in \cite{GSS} for a short proof).

\medskip

\begin{remark}\label{right-leftcont remark}
\cite{GSS} consider the left-continuous modification of admissible strategies. Since the respective formulas \eqref{ModelEquation} and \eqref{BV costs} for the price process and the costs of a strategy $X\in \cXBV(x,T)$ depend only on the measure $dX_t$, it is just a matter of notational convention whether to choose the right- or left-continuous modification of $X$. In particular, our formulas for the price process $S^X$ and the costs $\cC(X)$ are the same as those in \cite{GSS}.
Later on, however, we will consider a larger class of semimartingale strategies, and since  semimartingales are right-continuous by default and for good reason, we must adopt the convention of right continuity so as to be consistent between our two classes of strategies.

As can be seen from the formula \eqref{ow solution}, optimal strategies will typically have jumps at times $t=0$ and $t=T$. For right-continuous strategies, we need to include the possibility of an initial jump by allowing for an initial value $X_{0-}$ that can be different from $X_0$. Similarly, for the left-continuous modification of strategies used in \cite{GSS}, the terminal jump must be accommodated by allowing for a nonzero value of $X_T$ and by requiring the modified liquidation constraint $X_{T+}=0$. So both conventions require us  to impose conditions on the limits of $X_t$ when $t$ approaches a boundary point of the actual trading interval $[0,T]$ from outside this interval.  \hfill$\diamondsuit$
\end{remark}

\medskip

Here, our goal is to study the minimization of the expected costs  $\bE[\,\cC(X)\,]$ when $S^0$ has an additional drift. This topic is of intrinsic mathematical interest, and we  refer to the introduction of this paper for an account of our economic motivation to study this problem.
We assume henceforth that $S^0$ is a c\`adl\`ag semimartingale with decomposition
\begin{equation}\label{}
S^0_t=S_{0}+M_t+A_t,
\end{equation}
 where $S_0$ is a constant, $M$ is a square-integrable c\`adl\`ag martingale with $M_0=0$, and  $A$ is an  adapted process with $A_0=0$ and locally square-integrable total  variation, i.e., for every $T>0$ we have $\bE[\,|A|_{[0,T]}^2\,]<\infty$ when $|A|_{[0,T]}$ denotes the total variation of $A$ over the interval $[0,T]$.  There is in fact no loss of generality in assuming that $A$ is predictable (see Proposition I.4.23 in \cite{JacodShiryaev}).

It will turn out that the presence of $A$ increases the complexity of the optimal trade execution problem significantly.  In particular, optimal execution strategies in $\cXBV(x,T)$ will exist only under very restrictive assumptions on $A$. For instance, they will not exist even in the simple case in which $S^0$ is a diffusion model,
$$dS^0_t=\sigma(S^0_t)\,dW_t+b(S^0_t)\,dt,
$$
with nonconstant drift coefficient $b(\cdot)$. We therefore  need to extend our class of admissible trading strategies.

\begin{definition}\label{semimartingale strategy def} An \emph{admissible semimartingale strategy} is a  bounded\footnote{The requirement that $X$ is bounded is natural from an economic point of view, because the total number of available shares is finite for every stock. } right-continuous semimartingale $X$ for which  there exists a liquidation time $T\ge0$ such that $X_t=0$ $\bP$-a.s. for all $t\ge T$.
 By $\cXsem(x,T)$ we denote the class of all admissible semimartingale strategies $X$ with $X_{0-}=x$ and liquidation time $T$.
\end{definition}

\bigskip

Note that $\cXBV(x,T)$ is a subset of $\cXsem(x,T)$.
While semimartingale strategies are standard in frictionless asset pricing models, their application in a high-frequency market impact model is economically less natural than strategies of bounded variation, because they can no longer be written as the superposition of   buying and  selling strategies.

Given a semimartingale strategy $X\in\cXsem(x,t)$, we need to extend the definitions \eqref{ModelEquation} and \eqref{BV costs} for the corresponding price process and the resulting liquidation costs.  These formulas and our further analysis will involve stochastic integrals in which $X$ appears both as integrand and as integrator. Therefore, we first need to clarify how stochastic integrals must be understood in view of our requirement $X_{0-}=x\neq0$.

\bigskip

\begin{remark}[On the definition of stochastic integrals]\label{Stoch integrals remark} It is a common assumption in the literature on stochastic integration that semimartingales $X$ may jump at $t=0$, but a typical convention is to assume $X_{0-}=0$. With this convention, a stochastic integral $X_-\cdot Y$, as defined, e.g., in \cite{Protter}, will not depend on the initial jump of the integrator $Y$ at time $t=0$, and so there is no ambiguity in writing $(X_-\cdot Y)_t=\int_0^tX_{s-}\,dY_s$.
When the value $X_{0-}$ is nonzero, as it is the case for the semimartingale strategies defined above, one must carefully distinguish whether an initial jump of the integrator is or is not part of a stochastic integral. This has been done, e.g., by  \cite{Meyer}, from where we  adopt the convention of writing $\int_{[0,t]}X_{s-}\,dY_s$ or $\int_{(0,t]}X_{s-}\,dY_s$, respectively, when the initial jump is or is not part of the stochastic integral. We then have
\begin{equation}\label{stochastic integral convention}
\int_{[0,t]}X_{s-}\,dY_s=X_{0-}\Delta Y_0+\int_{(0,t]}X_{s-}\,dY_s\qquad\text{and}\qquad [X,Y]_0=\Delta X_0\Delta Y_0.
\end{equation}
 The integration by parts formula for stochastic integrals becomes
\begin{equation}\label{integration by parts}
X_tY_t=X_{0-}Y_{0-}+\int_{[0,t]}X_{s-}\,dY_s+\int_{[0,t]}Y_{s-}\,dX_s+[X,Y]_t
\end{equation}
see \cite[p. 303]{Meyer}. When $Z_t:=\int_{[0,t]}X_{s-}\,dY_s$ is a stochastic integral, we set $Z_{0-}:=0$ by default.\hfill$\diamondsuit$
\end{remark}

\bigskip

Given a semimartingale strategy $X\in\cXsem(x,T)$, the price $S^X_t$ at time $t$ can be defined just as in \eqref{ModelEquation} when $\int_{[0,t)}e^{-\rho(t-s)}\,dX_s$ denotes the left-hand limit, $E^X_{t-}$, of the generalized Ornstein-Uhlenbeck  process
\begin{equation}\label{E def}
E^X_t:=e^{-\rho t}\int_{[0,t]}e^{\rho s}\,dX_s,\qquad t\ge0.
\end{equation}

We now turn to the definition of the liquidation costs of the semimartingale strategy $X$. We will motivate our definition by an approximation from the discrete-time case. To this end, we take $N\in\bN$, let $t^N_k:=kT/N$ for $k=0,\dots, N$ and define the following sequence  of discrete trades:
$$\xi^N_0:=X_0-X_{0-}\qquad\text{and, for $k=1,\dots, N$,}\qquad \xi_k^N := X_{t_k^N} - X_{t_{k-1}^N}. $$
 Then, $\bm\xi^N:=(\xi^N_k)$ is an admissible trading strategy in the sense of \cite{ASS}. In Proposition 1 of \cite{ASS} and its proof, the costs incurred by  the discrete-time strategy $\bm\xi^N$ were derived as
\begin{align*}
	\cC^N(\bm\xi^N) =  \sum_{k=0}^N \Big(S^0_{t_k^N} \xi^N_k +  \sum_{i=0}^{k-1}e^{-\rho(t_k^N-t_i^N)}\xi_i^N \xi^N_k + \frac12  (\xi^N_k)^2\Big).
\end{align*}
The economic motivation of this formula is analogous to the one given in Remark~\ref{cost remark}. In fact $\cC^N(\bm\xi^N)$ coincides with $\cC(X^N)$, when $X^N\in \cXBV(x,T)$ denotes the  step function with jumps described by $\bm\xi^N$. We have the following asymptotics of these costs when our time grid becomes finer.

\begin{lemma}[Liquidation costs of a semimartingale strategy]\label{SemimartingaleCostLemma}As $N\ua\infty$, we have
$$\cC^N(\bm\xi^N)\longrightarrow  \int_{[0,T]} S^0_{t-} \,dX_t + [S^0,X]_T+\int_{[0,T]} E^X_{t-} \,dX_t + \frac12 [X]_T=:\cC(X)
$$
in probability, where $\cC(X)$ is independent of the (arbitrary) choice of the value $S_{0-}^0$, and $E^X$ is the generalized Ornstein-Uhlenbeck process from \eqref{E def}.\end{lemma}

\bigskip

We therefore define $\cC(X)$ as the \emph{liquidation costs} incurred by $X\in\cXsem(x,T)$. Note that $\cC(X)$ reduces to the liquidation costs defined in \eqref{BV costs} when $X\in\cXBV(x,T)$. Moreover, it follows  from \eqref{stochastic integral convention} that $\cC(X)$ is indeed independent of the particular choice of $S_{0-}^0$.

\subsection{Minimizing the expected costs}

The optimization problem we are interested in  is the minimization of the expected costs,
\begin{align} \label{MinProblemReduced}
  \bE[\,\cC(X)\,]=   \bE\left[\, \int_{[0,T]} S^0_{t-} \,dX_t + [S^0,X]_T + \int_{[0,T]}  E^X_{t-} \,dX_t + \frac12 [X]_T \,\right] ,
\end{align}
over all strategies $X$ that belong to $\cXsem(x,T)$ or to $\cXBV(x,T)$.
To state its solution, let
$Z=(Z_t)$ be a c\`adl\`ag version of the martingale
\begin{align*}
-	\bE\Big[\, A_T + \rho \int_0^T A_s \,ds\,\Big|\,\cF_t\,\Big],
\end{align*}
which exists due to our assumption that $(\Om,\cF,(\cF_t),\bP)$ satisfies the usual conditions. We also define the semimartingale $Y$ as
\begin{align*}
	Y_t :=  Z_t + \rho \int_0^t A_s\,ds +\big(1 + \rho(T-t)\big)A_t.
\end{align*}

\bigskip

\begin{theorem} \label{GenSemThm} When $A$ is $\bP$-a.s. absolutely continuous on $[0,T)$ with square-integrable derivative $A'_t=dA_t/dt$, i.e., when $A_t=\int_0^tA'_s\,ds$ for $0\le t<T$ and $\bE[\,\int_0^T(A'_t)^2\,dt\,]<\infty$, then
\begin{eqnarray}\label{OptimalCostSemimartingaleCaseEq}
\inf_{X\in\cXBV(x,T)} \bE \left[\, \cC(X)\, \right]&=&\inf_{X\in\cXsem(x,T)} \bE \left[\, \cC(X)\, \right]\\ &=& -xS_0 +\frac{x^2}{2+ \rho T} + \frac{xY_0}{2+\rho T} - \frac\rho{4}\bE \left[\,  \int_0^T \left( \frac{Y_s}{2+\rho(T-s)} - \frac1\rho A'_s\right)^2 \,ds \,\right],\nonumber
\end{eqnarray}
and
\begin{equation}\label{infinite cost}
\inf_{X\in\cXBV(x,T)} \bE \left[\, \cC(X)\, \right]=\inf_{X\in\cXsem(x,T)} \bE \left[\, \cC(X)\, \right] = -\infty
\end{equation}
otherwise.

When in addition $A'$ is bounded,  the second infimum in \eqref{OptimalCostSemimartingaleCaseEq} can be attained  only if $A'$ is a (right-continuous) semimartingale, and the unique optimal strategy is then given by
\begin{equation}\label{X optimal strategy}
\begin{split}
X_t&=\frac{x(1+\rho( T-t))-\frac{1}2 (1+\rho t)Y_0}{2+\rho T}-\frac12\int_{(0,t]}\varphi(s)\,dZ_s+\frac1{2\rho}A'_t\\
&\qquad\qquad\qquad\qquad-\rho\int_0^t\bigg(\frac12\int_{(0,s]}\varphi(r)\,dZ_r+\frac12A_s\bigg)\,ds,
\end{split}
\end{equation}
where $ \varphi(t) := (2+\rho(T-t))^{-1}$.
In particular the first infimum in \eqref{OptimalCostSemimartingaleCaseEq} can only be attained when  $\frac1\rho A'_t-\int_{(0,t]}\varphi(s)\,dZ_s$
is $\bP$-a.s. right-continuous and of finite variation on $[0,T]$.\end{theorem}

$ $

\begin{remark} From an economic point of view, the fact that it is possible to generate arbitrarily negative expected costs for drift processes that are not absolutely continuous might indicate a market inefficiency that arises when trading takes place on a much shorter time scale than the resilience of price impact. The market then becomes inefficient, because its resilient reaction to a price shock is delayed in comparison to the trading activities of the economic agent; see also Remarks 2 and 3 in \cite{ASS}. This becomes particularly apparent when the drift is generated by the trading behavior of a large fundamental seller, who is subject to predatory trading by a high-frequency trader; see Remark~\ref{TwoplayerRemark} below. 

We refer to Lemma~\ref{infinite cost when not abs Lemma} in Section~\ref{Proofssection} for the details of constructing a strategy with arbitrarily negative expected costs when $A$ is not absolutely continuous.\hfill$\diamondsuit$
\end{remark}

\medskip

The situation in Theorem~\ref{GenSemThm} simplifies significantly when $A'$ is a martingale:

\medskip

\begin{corollary}\label{A0 martingale corollary} Suppose that $A$ is of the form $A_t=\int_0^tA'_s\,ds$ for a bounded c\`adl\`ag martingale $A'$. Then the optimal strategy \eqref{X optimal strategy} becomes
\begin{equation}\label{A0 martingale optimal strategy}
   X_t = \frac{x(1+\rho(T-t))}{2+\rho T} + \frac1{4\rho}(2+\rho(T-t))A'_t + \frac14(1+\rho(T-t))A_t .
\end{equation}
\end{corollary}

\medskip

\begin{figure}[htbp]
\begin{center}
\includegraphics[width=8cm]{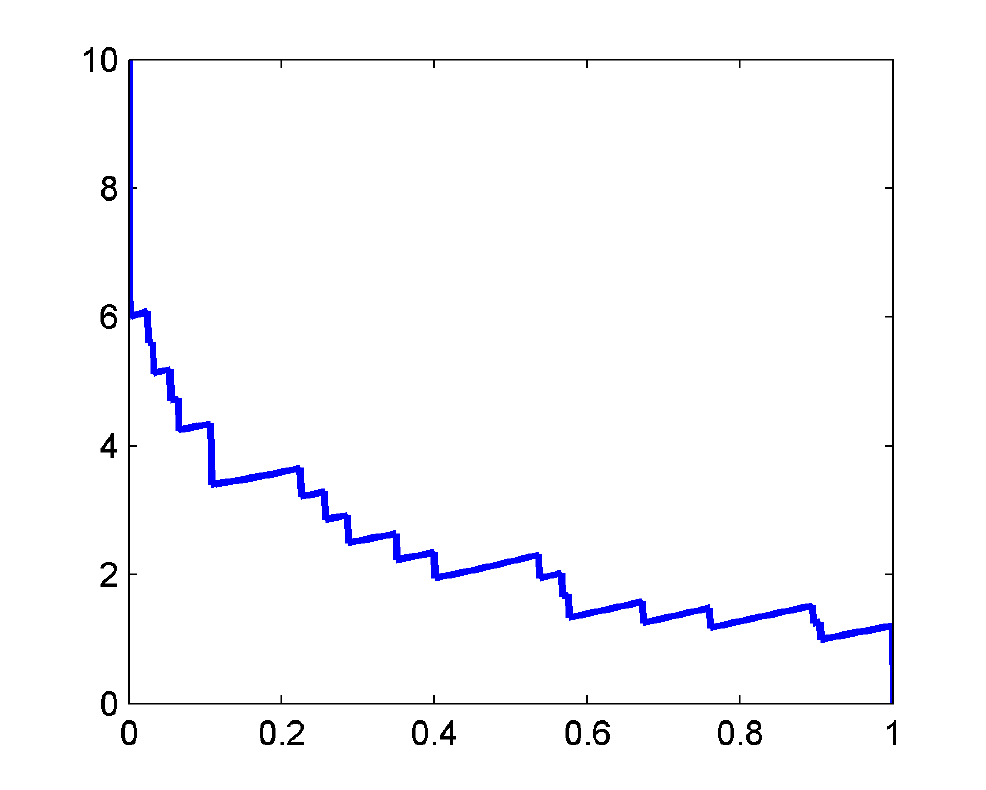}
\caption{Optimal strategy \eqref{A0 martingale optimal strategy} when $\rho=2$ and $A'$ is a compensated Poisson  process with intensity $\lambda=20$. }\label{jumpfigure}
\end{center}
\end{figure}

Note that the  strategy \eqref{A0 martingale optimal strategy} can be computed in a pathwise manner without reference to the particular distribution of $A$; see Figure~\ref{jumpfigure}. This special case highlights the ambiguous and seemingly contradictory nature of the robustness of the optimal strategy: this strategy reacts very sensitively to structural features of the price process, i.e., to the martingale property of $A'$, but once this structural requirement is satisfied, the strategy is completely independent of the law of $A'$. When $A$ vanishes, this strategy reduces to the Obizhaeva--Wang solution \eqref{ow solution}.

\begin{remark}[Comparison with Almgren--Chriss model]It is interesting to compare the optimal strategy \eqref{OptimalCostSemimartingaleCaseEq} with the one for the corresponding Almgren--Chriss model. In the latter model, strategies must be absolutely continuous. Given such a strategy $X$,  the price process takes the form
$$\wt S^X_t=S^0_t+\eta\dot X_t+{\gamma (X_t-X_0)},
$$
where $\eta$ and $\gamma$ are two nonnegative constants. When $X_0=x$ and $X_T=0$, the corresponding liquidation costs are
$$\wt\cC(X)=\int_0^TS^0_t\dot X_t\,dt+\eta\int_0^T\dot X_t^2\,dt+\frac\gamma2x^2.
$$
In our setting, there is always a  unique strategy that minimizes the expected liquidation costs $\bE[\,\wt\cC(X)\,]$ and it is given by
$$X_t=\frac{T-t}{T}\bigg(x-\frac1{2\eta}\int_0^t\frac T{(T-s)^2}\bE\Big[\,\int_s^T(T-u)\,dA_u\,\Big|\,\cF_s\,\Big]\,ds\bigg);
$$
see Corollary 2 in \cite{SchiedRobust}. Here the drift $A$ enters the optimal strategy  basically in integrated form, and so one can expect that  possible misspecifications of the drift may average out to some extent.  This relatively stable behavior should be compared to the direct dependence of the  strategy \eqref{OptimalCostSemimartingaleCaseEq} on the {derivative} of the drift.\hfill$\diamondsuit$
\end{remark}

\begin{remark}[A two-player situation] \label{TwoplayerRemark}
As discussed in the Introduction, an important source for a drift in the asset price process $S^0$ can be the trading activity of another large market participant (\lq\lq the seller"). There are various reasons why another economic agent (\lq\lq the predator") may get good estimates for the resulting drift. For instance, some trade execution algorithms create characteristic order patterns and therefore allow for an inference of their future trading trajectory. But there are also other possibilities as discussed in \cite{SchoenebornSchied}. 

Suppose that the seller aims at  liquidating a position of $x\neq0$ shares by time $T>0$. Suppose moreover, for simplicity, that  the unaffected asset price $S^0$ is a square-integrable martingale so that the seller will use the liquidation strategy $X^*$ from \eqref{ow solution}. The predator will then perceive the unaffected price process $\wt S^0=S^0+E^{X^*}$, which is no longer a martingale but has the drift $E^{X^*}$.
   Since $X^*$ has a terminal jump, also the resulting \lq drift\rq\ $E^{X^*}$ will jump by the same amount at time $T$.  So if the predator faces a more relaxed time constraint than the seller, which is a   natural assumption, the predator will perceive a drift that is not absolutely continuous and, by Theorem~\ref{GenSemThm}, will have the possibility of making arbitrary large expected profits. Similar results will also hold when $S^0$ has a nonvanishing drift.\hfill$\diamondsuit$
\end{remark}

\subsection{Minimization of a cost-risk criterion}

 As a corollary to Theorem~\ref{GenSemThm}, we can  also find optimal strategies for the linear risk criterion that was proposed in \cite{GatheralSchied} for the  Almgren--Chriss framework with a risk-neutral geometric Brownian motion as unaffected price process.  When in our model $S^0$ is a risk-neutral geometric Brownian motion, i.e.,
\begin{equation}\label{eq:GBM}
S^0_t=S_0e^{\sigma\, W_t-\frac12\,\sigma^2\,t}\qquad\text{for some $\sigma>0$ and a Brownian motion $W$,}
\end{equation}
the same reasoning as in \cite{GatheralSchied} motivates the minimization of a cost-risk functional of the form
\begin{equation}\label{GS risk functional}
\bE\Big[\,\cC(X)+\lambda\int_0^TS^X_tX_t\,dt\,\Big],
\end{equation}
where  $\lambda$ has the same sign as $X_{0-}=x$. The parameter $\lambda$ is typically derived from the Value at Risk of a unit asset position under the assumption of  log-normal future returns. As argued in Remark 2.2 of \cite{GatheralSchied}, one could obtain the same cost-risk functional (but perhaps with a different value for $\lambda$) if Value at Risk is replaced by a coherent risk measure or by any other positively homogeneous risk measure.

Optimal strategies for the cost-risk functional \eqref{GS risk functional} in the Almgren--Chriss framework have the advantages of being sensitive to changes in the asset price,  easily computable in closed form, and possess completely transparent reactions to parameter changes. In addition, they have a striking robustness property: they are independent of the actual law of $S^0$ as long as $S^0$ is a martingale. Thus they may be optimal even when the law of $S^0$ is not of the particular form   \eqref{eq:GBM}. A disadvantage is that optimal strategies can switch sign, in which case  the interpretation of the cost-risk functional \eqref{GS risk functional} breaks down. But,   as discussed in Section 4 of \cite{GatheralSchied}, the probability that strategies  become negative will be small with reasonable parameter choices.

\bigskip

\begin{corollary}\label{cost-risk corollary}
The minimization of the cost-risk functional \eqref{GS risk functional} is equivalent to the minimization of the expected costs for the new price process
$$\wt S^0_t =\frac\rho{\rho+\lambda}\bigg(S^0_t-\lambda\int_0^tS^0_s\,ds\bigg).
$$
In particular, the statements of Theorem~\ref{GenSemThm}  carry over to the minimization of the cost-risk functional \eqref{GS risk functional} when $A$ is replaced by
$$\wt A_t=\frac\rho{\rho+\lambda}\bigg(A_t-\lambda\int_0^tS^0_s\,ds\bigg).
$$
\end{corollary}

\bigskip

When $S^0$ is a {bounded} martingale, then $\wt A'_t=-\frac{\rho\lambda}{\rho+\lambda}S^0_t$ is also a martingale.  Thus, by Corollary~\ref{A0 martingale corollary} the optimal strategy that minimizes the cost-risk criterion \eqref{eq:GBM} simply becomes
$$
X^*_t=\frac{x(1+\rho( T-t))}{2+\rho T} -\frac{\rho\lambda}{\rho+\lambda}\bigg( \frac1{4\rho}(2-\rho(T-t))S^0_t \ + \ \frac14(1+\rho(T-t))\int_0^tS^0_s\,ds\bigg).
$$
This strategy can be computed in a pathwise manner and is completely independent of the particular law of the martingale $S^0$. It thus minimizes the cost-risk criterion \eqref{eq:GBM} whenever $\bP$ is a martingale measure for $S^0$. When $S^0$ is not bounded but  just a square-integrable martingale, then $X^*$ will not be an admissible semimartingale strategy in the sense of Definition~\ref{semimartingale strategy def}. Nevertheless,
in this special case, one can show that  $X^*$ attains the optimum of the cost-risk criterion and thus can still be regarded as an optimal strategy. We leave the details to the reader.

\section{Proofs}\label{Proofssection}

To simplify the notation,  we will drop the superscript $X$ in $E^X$ throughout the proofs when there is no ambiguity about the strategy $X$ used in the definition of $E=E^X$.

\bigskip\noindent{\bf Proof of Lemma~\ref{SemimartingaleCostLemma}.} We first note that
\begin{align*}
\sum_{k=0}^N S^0_{t_k^N} \xi^N_k &= S_0^0\Delta X_0+\sum_{k=1}^N S^0_{t_{k-1}^N} (X_{t_k^N} - X_{t_{k-1}^N}) + \sum_{k=1}^N (S^0_{t_k^N}-S^0_{t_{k-1}^N}) (X_{t_k^N} - X_{t_{k-1}^N}).
\end{align*}
By Theorems II.5.21 and II.5.23 in \cite{Protter}, this expression converges in probability to
\begin{eqnarray*}
S^0_0\Delta X_0+ \int_{(0,T]} S^0_{t-} \,dX_t + [S^0,X]_T -\Delta S^0_0\Delta X_0=\int_{[0,T]} S^0_{t-} \,dX_t + [S^0,X]_T.
\end{eqnarray*}
Similarly,\begin{align*}
	\sum_{k=0}^N (\xi_k^N)^2 =(\Delta X_0)^2 +\sum_{k=1}^N (X_{t_k^N} - X_{t_{k-1}^N})^2 \longrightarrow [X]_T,
\end{align*}
in probability.

 When defining
$$\wt E_t^N := \sum_{i=0}^{N-1} e^{\rho t_i^N} (X_{t_{i+1}^N \wedge t} - X_{t_i^N \wedge t})
$$
then $\wt E^N_t$ is the Riemann approximation  of a stochastic integral with a deterministic and continuous integrand, and hence
$\wt E^N \rightarrow \int_{(0,\cdot]}e^{\rho s}\,dX_s$
 uniformly  on compacts in probability (ucp) \cite[Proposition I.4.44]{JacodShiryaev}. It follows that
$$E_t^N:=e^{-\rho t}\big(e^{-\rho T/N}\Delta X_0+\wt E^N_t\big)\longrightarrow E_t\qquad \text{ucp as $N\ua\infty$.}$$

Moreover,
\begin{eqnarray*}\sum_{i=0}^{k-1}e^{-\rho(t_k^N-t_i^N)}\xi_i^N& =&e^{-\rho t_k^N}\Big(\Delta X_0+\sum_{i=1}^{k-1}e^{\rho t_i^N}(X_{t_{i}^N}-X_{t_{i-1}^N})\Big)\\
&=&e^{-\rho t_{k-1}^N}\Big(e^{-\rho T/N}\Delta X_0+\sum_{i=0}^{k-2}e^{\rho t_i^N}(X_{t_{i+1}^N}-X_{t_{i}^N})\Big)=E^N_{t_{k-1}^N}.
\end{eqnarray*}
Therefore,
\begin{align*}
\sum_{k=0}^N	\sum_{i=0}^{k-1}e^{-\rho(t_k^N-t_i^N)}\xi_i^N \xi^N_k &=\sum_{k=1}^N	 \sum_{i=0}^{k-1}e^{-\rho(t_k^N-t_i^N)}\xi_i^N \xi^N_k= \sum_{k=1}^N E_{t^N_{k-1}}^N (X_{t_k^N} - X_{t^N_{k-1}})\\
& =\int_{(0,T]} (E^N)^{\sigma_N}_t\,dX_t,
\end{align*}
where, using the notation from Section II.5 of \cite{Protter}, for a process $Y$ we let
$$Y^{\sigma_N}:=Y_0\eins_{\{0\}}+\sum_{k=0}^{N-1}Y_{t_k^N}\eins_{(t_k^N,t_{k+1}^N]}\qquad\text{and}\qquad \int_{(0,T]} Y^{\sigma_N}_t\,dX_t=\sum_{k=0}^{N-1}Y_{t_k^N}(X_{t_{k+1}^N}-X_{t_k^N}).
$$
Now
\begin{eqnarray}\int_{(0,T]}(E^N)^{\sigma_N}_t\,dX_t=\int_{(0,T]}E^{\sigma_N}_t\,dX_t+\int_{(0,T]}\Big((E^N)^{\sigma_N}_t-E^{\sigma_N}_t\Big)\,dX_t.\label{E^N to E auxiliary eq}
\end{eqnarray}
The first integral on the right converges  to $\int_{(0,T]}E_{t-}\,dX_t$ in probability by Theorem II.5.21 of \cite{Protter}. To deal with the second integral on the right, we note that $\sup_{0\le t\le T}|E^N_t-E_t|\le\eps$ implies that also  $\sup_{0\le t\le T}|(E^N)^{\sigma_N}_t-E_t^{\sigma_N}|\le\eps$. Thus,  $(E^N)^{\sigma_N}-E^{\sigma_N}\to0 $ ucp.  The continuity of the stochastic integral with respect to ucp convergence \cite[p. 59]{Protter} therefore implies that the rightmost integral in \eqref{E^N to E auxiliary eq} tends to zero in probability for $N\ua\infty$. We thus obtain that
$$\sum_{k=0}^N	\sum_{i=0}^{k-1}e^{-\rho(t_k^N-t_i^N)}\xi_i^N \xi^N_k\longrightarrow \int_{(0,T]}E_{t-}\,dX_t=\int_{[0,T]}E_{t-}\,dX_t
$$
in probability (here we have used the fact that $E_{0-}=0$ by our convention on stochastic integrals made at the end of Remark~\ref{Stoch integrals remark}).
 Putting everything together yields the assertion. \qed

\bigskip

Now we start preparing for the proof of Theorem~\ref{GenSemThm}, which will rely on a series of lemmas.  The basic idea underlying the proof is the verification argument  appearing in the next lemma.  The nature of the verification argument becomes apparent when taking $\alpha_t:=A'_t$ in Lemma~\ref{CostLemma}. The key to the argument is the  following formula for the  remaining costs of  optimally liquidating the asset position $X_t$ over $(t,T]$, taking into account a given volume impact $E_t$. This volume impact $E_t$ can be thought of as the volume impact generated by using a strategy $X$ throughout $[0,t]$ that leads to the asset position $X_t$ at time $t$. The formula is
\begin{equation}\label{costs guess}
- \frac12 E_t^2+ \varphi(t)(X_t-E_t)^2 +\varphi(t)(X_t-E_t)Y_t -\rho\bE\bigg[\,\int_t^T\Big(\frac12\varphi(s)Y_s-\frac1{2\rho}A'_s\Big)^2\,ds\,\Big|\,\cF_t\,\bigg].
\end{equation}
 This formula needs to be guessed; we are not aware of a method by which it can be derived analytically. Once this formula has been guessed, we can proceed by the following standard verification argument, which is also used, e.g., in Section 6.6.1 of \cite{Pham}: We show that the costs \eqref{costs guess} plus the costs generated by using $X$ over $[0,t]$ is submartingale for any strategy $X$ and a true martingale if $X$ is an optimal strategy.

Let us  recall the definition
\begin{align*}
    \varphi(t) = \frac{1}{2+\rho(T-t)}.
\end{align*}

\medskip

\begin{lemma}\label{CostLemma}Fix $X\in\cXsem(x,T)$, and let $\alpha_t$ be any progressively measurable process with $\bE[\,\int_0^T\alpha_t^2\,dt\,]<\infty$. We furthermore let $Z^\alpha = (Z^\alpha_t)$ be a c\`adl\`ag version of the martingale
$$ - \bE\bigg[\,\int_0^T\alpha_s\,ds+\rho \int_0^T\int_0^s \alpha_r \,dr\,ds\,\Big|\,\cF_t\,\bigg],
$$
 and we define
\begin{align*}
	Y^\alpha_t :=  Z^\alpha_t + \rho \int_0^t \int_0^s\alpha_r\,dr\,ds +\big(1+ \rho(T-t)\big)\int_0^t\alpha_s\,ds.
\end{align*}
Then
\begin{equation}\label{ExpectedCostsAlphaEq}\begin{split}
\bE \left[\, \cC(X)\, \right]&=-xS_0+\varphi(0){x^2}  +\varphi(0){xZ^\alpha_0} - \rho\bE \left[\,  \int_0^T \left(\frac12 \varphi(s){Y^\alpha_s}- \frac1{2\rho} \alpha_s\right)^2 \,ds \,\right]\\
&\quad+\bE\Big[\,\int_0^TX_t\alpha_t\,dt-\int_{(0,T]}X_{t-}\,dA_t\,\Big]\\
&\quad +\rho\bE\bigg[\int_0^{T}\bigg\{\varphi(t)X_t+(1-\varphi(t))E_t+\frac12\varphi(t)Y^\alpha_t-\frac1{2\rho}\alpha_t\bigg\}^2\,dt\,\bigg].
\end{split}
\end{equation}
\end{lemma}
\bigskip

\Proof
 We note first that 
  Jensen's inequality implies $(\int_0^T\alpha_t\,dt)^2\le T\int_0^T\alpha_t^2\,dt$. Hence,
\begin{equation}\label{Lemma 2 inequality}
\bE[\,(Z_T^\alpha)^2\,]=\bE\bigg[\,\bigg(\int_0^T\alpha_t\,dt+\rho \int_0^T\int_0^t \alpha_s \,ds\,dt\bigg)^2\,\bigg]\le\bE\bigg[\,\bigg((1+\rho T)\sqrt{T\int_0^T \alpha_s^2\,ds}\bigg)^2\,\bigg]<\infty,
\end{equation}
and in turn   $\bE[\,\int_0^T(Y_t^\alpha)^2\,dt]<\infty$. So all expressions in \eqref{ExpectedCostsAlphaEq} are well-defined.
We now define for $X\in\cXsem(x,T)$
$$
	\wt C^X_t := \int_{[0,t]} S^0_{t-} \,dX_t + [S^0,X]_t + \int_{[0,t]} E_{s-} \,dX_s + \frac12 [X]_t.$$
 Then $\wt C^X_t $ describes the costs incurred by using the strategy $X$ throughout the time interval $[0,t]$. Next, we use our guess \eqref{costs guess} for the  costs of optimally liquidating the amount $x=X_t$ by trading over $(t,T]$ when an initial volume impact of size $\eps=E_t$ is given at time $t$. It leads to defining the function
$$
	V^\alpha(t,x,\eps):=- \frac12 \eps^2+ \varphi(t)(x-\eps)^2 +\varphi(t)(x-\eps)Y^\alpha_t,
$$
which describes these optimal costs less the integral term in \eqref{costs guess}, which does not depend on $x=X_t$ or $\eps=E_t$. By adding $\wt C^X_t$ we get the process
\begin{equation}\label{cost formula}
C^X_t:=\wt C^X_t+V^\alpha(t,X_t,E_t).
\end{equation}

We will now compute the It\^o differential $dC^X_t$. Our computation will mainly rely on It\^o's product rule in the form \eqref{integration by parts}. For the computation, it will be helpful to collect a few auxiliary formulas in advance. For instance,  it follows from the definition of $Y^\alpha$ that
\begin{align}
	dY^\alpha_t = dZ^\alpha_t +(1+\rho(T-t)) \alpha_t\,dt  = dZ^\alpha_t + \frac{1-\varphi(t)}{\varphi(t)} \alpha_t\,dt .\label{Yalpha dynamics}
\end{align}
Using  $E_t=e^{-\rho t} \int_{[0,t]} e^{\rho s}\,dX_s$ from \eqref{E def}, the fact that  $E_{0-}=0$ (which follows from our corresponding convention for stochastic integrals), and integration by parts yields
\begin{equation}\label{EtSDE}
E_t = X_t -x- \rho \int_0^tE_s \,ds, \qquad 0\le t\le T.
\end{equation}
 It follows in particular that the process $E_t-X_t$ does not jump throughout $[0,T]$ and that on this interval $d(E_t-X_t)=-\rho E_t\,dt$.
 We also note that $d\varphi(t)=\rho\varphi(t)^2\,dt$.

 Recalling  the fact that $M_0=A_0=0$, we now choose values $S^0_{0-}$, $M_{0-}$, and $A_{0-}$ in $\bR$ that satisfy
\begin{align}\label{S0-}
   S_0- S^0_{0-}=\Delta S^0_0=\Delta M_0+\Delta A_0=-M_{0-}-A_{0-}
\end{align}
but can otherwise be arbitrary.  We also choose an arbitrary value $Z^\alpha_{0-}\in\bR$.
We then have on $[0,T]$
\begin{equation}\label{wt CX integration by parts}
\begin{split}d\wt C^X_t&=S^0_{t-}\,dX_t+d[S^0,X]_t+E_{t-}\,dX_t+\frac12\,d[X]_t\\
&=d(S^0_tX_t)-X_{t-}\,dM_t-X_{t-}\,dA_t+E_{t-}\,dX_t+\frac12\,d[X]_t.
\end{split}\end{equation}
Hence, a lengthy but straightforward calculation gives
\begin{eqnarray}{dC^X_t}&=&d(S^0_tX_t)-X_{t-}\,dM_t-X_{t-}\,dA_t+\varphi(t)(X_{t-}-E_{t-})\,dZ^\alpha_t\nonumber\\
&&\quad+\rho\bigg\{E_t^2+\varphi(t)^2(X_t-E_t)^2+2\varphi(t)(X_t-E_t)E_t\nonumber\\
&&\quad+\frac1\rho(1-\varphi(t))(X_{t}-E_{t})\alpha_t+\varphi(t)E_tY^\alpha_t+\varphi(t)^2(X_t-E_t)Y^\alpha_t \bigg\}\,dt\nonumber\\
&&\qquad\nonumber\\
&=&d(S^0_tX_t)-X_{t-}\,dM_t+\varphi(t)(X_{t-}-E_{t-})\,dZ^\alpha_t\\
&&\quad+\rho\bigg\{\varphi(t)X_t+(1-\varphi(t))E_t+\frac12\varphi(t)Y^\alpha_t-\frac1{2\rho}\alpha_t\bigg\}^2\,dt\label{dCx equation}\nonumber\\
&&\quad +X_{t-}(\alpha_t\,dt-dA_t)-\rho\Big(\frac12\varphi(t)Y^\alpha_t-\frac1{2\rho}\alpha_t\Big)^2\,dt.\nonumber
\end{eqnarray}
Due to the regularity of their sample paths, all stochastic processes involved have $\bP$-a.s. at most countably many discontinuities. The set of jump times therefore has Lebesgue measure zero. Hence we can replace left-hand limits in terms such as $\alpha_{t-}\,dt$ by their regular values, i.e., we can write $\alpha_t\,dt$. Moreover,  by \eqref{S0-} and the fact that $X_T=0$,
\begin{equation}\label{x0S0 lost eqn}\begin{split}
\int_{[0,T]}d(S^0_tX_t)-\int_{[0,T]}X_{t-}\,dM_t&=S^0_TX_T-S^0_{0-}X_{0-}-X_{0-}\Delta M_0-\int_{(0,T]}X_{t-}\,dM_t\\
&=-xS_0+X_{0-}\Delta A_{0}-\int_{(0,T]}X_{t-}\,dM_t.
\end{split}
\end{equation}
Building the integral $ \int_{[0,T]}dC^X_t$ thus  yields
\begin{eqnarray}C^X_T-C_{0-}^X&=&-xS_0-\int_{(0,T]}X_{t-}\,dM_t+\int_{[0,T]}\varphi(t)(X_{t-}-E_{t-})\,dZ^\alpha_t+\int_0^T                                                                                                                                                      X_t\alpha_t\,dt-\int_{(0,T]}X_{t-}\,dA_t\nonumber\\
&&+\rho\int_0^T\bigg\{\varphi(t)X_t+(1-\varphi(t))E_t+\frac12\varphi(t)Y^\alpha_t-\frac1{2\rho}\alpha_t\bigg\}^2\,dt\label{CXT-CX0}\\
&&-\rho\int_0^T\Big(\frac12\varphi(t)Y^\alpha_t-\frac1{2\rho}\alpha_t\Big)^2\,dt.\nonumber
\end{eqnarray}

The stochastic integral $L_u:=\int_{(0,u]}X_{t-}\,dM_t$ satisfies $\bE[\,[L]_T\,]=\bE[\,\int_{(0,T]}X_{t-}^2\,d[M]_t\,]<\infty$   since  $M$ is a square-integrable martingale  and $X$ is bounded by definition. Hence $L$ is a true martingale  and satisfies $\bE[\,L_T\,]=L_0=0$.

 Now we show that
\begin{equation}\label{Z stoch integral expectation}
\bE\Big[\,\int_{[0,T]}\varphi(t)(X_{t-}-E_{t-})\,dZ^\alpha_t\,\Big]=\varphi(0)x( Z^\alpha_{0}-Z^\alpha_{0-}).
\end{equation}
   Taking expectations in \eqref{CXT-CX0} will then yield the assertion, because, on the one hand, $\wt C_{0-}^X=0$, $Y^\alpha_{0-}=Z_{0-}^\alpha$, and so
$C^X_{0-}=V(t-,X_{0-},E_{0-})=\varphi(0)x^2+\varphi(0)xZ_{0-}^\alpha$. On the other hand,
\begin{equation}\label{YT=0}
Y^\alpha_T=0\qquad\text{$\bP$-a.s.}
\end{equation}
and so $C_T^X=\wt C^X_T+V^\alpha(T,0,E_T)=\cC(X)$ $\bP$-a.s.

To show \eqref{Z stoch integral expectation}, we use \eqref{Lemma 2 inequality} and Doob's quadratic maximal inequality to conclude that $Z^\alpha$ is a square-integrable martingale. Moreover,  the boundedness of $X$, the identity \eqref{EtSDE},  and Gronwall's lemma yield that $E$ is bounded as well. Thus, the stochastic integral $N_u:=\int_{[0,u]}\varphi(t)(X_{t-}-E_{t-})\,dZ^\alpha_t$ is a true martingale.
Together with $N_0=\varphi(0)x\Delta Z^\alpha_0$ this shows \eqref{Z stoch integral expectation} and thus concludes the proof. \qed

\bigskip

\begin{remark}In the preceding proof, we have only used the facts that $Y^\alpha$ satisfies \eqref{Yalpha dynamics} for some square-integrable martingale $Z^\alpha$  and the identity \eqref{YT=0}. But these two identities already determine $Y^\alpha$ and $Z^\alpha$.
\end{remark}

\bigskip

In the next lemma, we derive an explicit formula for a strategy for which the last term in \eqref{ExpectedCostsAlphaEq} vanishes. When we can take $\alpha_t=A'_t$, this strategy will be the optimal strategy.

\bigskip

\begin{lemma}\label{alpha strategy lemma}Suppose that $\alpha$ is a bounded semimartingale  and that $Y^\alpha$ and $Z^\alpha$ are as in Lemma~\ref{CostLemma}. Then for all $x\in\bR$ and $T>0$ there exists a unique strategy $X\in\cXsem(x,T)$ such that
\begin{equation}\label{XoptimalitycondAlphaEq}
\varphi(t)X_t+(1-\varphi(t))E_t+\frac12\varphi(t)Y^\alpha_t-\frac1{2\rho}\alpha_t=0\qquad\text{for a.e. $t\in[0,T)$.}
\end{equation}
Moreover, for $0\le t<T$, $X$ is given by
\begin{equation}\label{Xalpha strategy}
\begin{split}
X_t&=\frac{x(1+\rho( T-t)) -\frac{1}2 (1+\rho t)Y_0^\alpha}{2+\rho T}-\frac12\int_{(0,t]}\varphi(s)\,dZ^\alpha_s+\frac1{2\rho}\alpha_t\\
&\qquad+\rho\int_0^t\bigg(-\frac12\int_{(0,s]}\varphi(r)\,dZ^\alpha_r-\frac12\int_0^s\alpha_r\,dr\bigg)\,ds.
\end{split}
\end{equation}
 Furthermore, $X$ has the initial jump
\begin{equation}\label{Xalpha initial jump}
 \Delta X_0=\frac1{2\rho}\alpha_0-\frac{x+\frac 12 Y_0^\alpha}{2+\rho T}
\end{equation}
and  the terminal jump
\begin{equation}\label{Xalpha terminal jump}
\Delta X_T=-\frac{x+\frac 12 Y_0^\alpha}{2+\rho T}-\frac12\int_{(0,T)}\varphi(s)\,dZ^\alpha_s-\frac12\int_0^T\alpha_s\,ds+\frac12Y^\alpha_{T-}-\frac1{2\rho}\alpha_{T-}.
\end{equation}
In particular, $X$ belongs to $\cXBV(x,T)$ when both $Z$ and $\alpha$ are of finite variation.
\end{lemma}

\Proof When \eqref{XoptimalitycondAlphaEq} is not already satisfied at $t=0-$, then an initial jump is needed
so that \eqref{XoptimalitycondAlphaEq} is satisfied immediately after the jump, because all processes in \eqref{XoptimalitycondAlphaEq} are right-continuous.  Taking the limit $t\da0$ in \eqref{XoptimalitycondAlphaEq} and using the identities $X_0=x+\Delta X_0$ and  $E_0=\Delta X_0$  yields
\begin{equation}\label{Delta X0 eq}
 \Delta X_0=-\frac12\varphi(0)Y_0^\alpha + \frac1{2\rho } \alpha_0 -\varphi(0)x=\frac1{2\rho}\alpha_0-\frac{x+\frac 12 Y_0^\alpha}{2+\rho T}.
\end{equation}

We now solve for the dynamics of $X$ on $(0,T)$.
 Dividing \eqref{XoptimalitycondAlphaEq} by $\varphi$ and taking differentials yields
\begin{eqnarray}
0&=&dX_t-\rho E_t\,dt+(1+\rho(T-t))\,dE_t+\frac12\,dY^\alpha_t+\frac12\alpha_t\,dt-\frac1{2\rho \varphi(t)}\,d\alpha_t\nonumber\\
&=&\frac1{\varphi(t)}\,dE_t+\frac12\,dY^\alpha_t+\frac12\alpha_t\,dt-\frac1{2\rho \varphi(t)}\,d\alpha_t,\label{dE equation}
\end{eqnarray}
where we have used the identity $dE_t=dX_t-\rho E_t\,dt$ in the second step. We can now informally solve \eqref{dE equation} for $dE_t$ and then obtain that for $t\in[0,T)$
\begin{equation}\label{E equation integrated}
E_t=\Delta X_0+\int_{(0,t]}dE_s=-\frac{x+\frac 12 Y^\alpha_0}{2+\rho T}-\frac12\int_{(0,t]}\varphi(s)\,dZ^\alpha_s-\frac12\int_0^t\alpha_s\,ds+\frac1{2\rho}\alpha_t,
\end{equation}
where we have used the fact that $E_0=\Delta X_0$, \eqref{Delta X0 eq}, and \eqref{Yalpha dynamics}. To make this argument rigorous, we define $\Lambda_t$ so that  \eqref{dE equation} becomes
$0=\int_{(0,t]}\frac1{\varphi(s)}\,dE_s+\Lambda_t$ after integration. Then we use the associativity of the stochastic integral \cite[Theorem II.5.19]{Protter} to get
$$0=\int_{(0,t]}\varphi(s)\frac1{\varphi(s)}\,dE_s+\int_{(0,t]}\varphi(s)\,d\Lambda_s=E_t-E_0+\int_{(0,t]}\varphi(s)\,d\Lambda_s.
$$
When taking differentials again, we arrive at \eqref{E equation integrated}.

Now \eqref{EtSDE} and \eqref{E equation integrated} yield that for $t\in[0,T)$
\begin{eqnarray*}X_t&=&x+E_t+\rho\int_0^tE_s\,ds\\
&=&\frac{x(1+\rho( T-t))-\frac{1}2 (1+\rho t)Y_0^\alpha}{2+\rho T}-\frac12\int_{(0,t]}\varphi(s)\,dZ^\alpha_s+\frac1{2\rho}\alpha_t\\
&&\qquad+\rho\int_0^t\bigg(-\frac12\int_{(0,s]}\varphi(r)\,dZ^\alpha_r-\frac12\int_0^s\alpha_r\,dr\bigg)\,ds
\end{eqnarray*}
and this proves \eqref{Xalpha strategy}. It is moreover clear from the proof that any strategy in $\cXsem(x,T)$ satisfying \eqref{XoptimalitycondAlphaEq} must be of this form, which gives uniqueness.

Now we turn to proving our formula for the terminal jump.  Taking left-hand limits $t\ua T$ in \eqref{XoptimalitycondAlphaEq} and using that $\varphi(T-)=\varphi(T)=1/2$ yields
\begin{eqnarray*}
0&=&X_{T-}+E_{T-}+\frac12Y^\alpha_{T-}-  \frac1{\rho}\alpha_{T-}\\
&=&X_{T-}-\frac{x+\frac 12 Y^\alpha_0}{2+\rho T}-\frac12\int_{(0,T)}\varphi(s)\,dZ^\alpha_s-\frac12\int_0^T\alpha_s\,ds+\frac12Y^\alpha_{T-}-\frac1{2\rho}\alpha_{T-},
\end{eqnarray*}
where we have used \eqref{E equation integrated} in the second step.
Since  $X_T=0$ we have $\Delta X_T=-X_{T-}$, and our formula follows.

Now we show that $X$ is admissible, i.e., we must show that $X$ is bounded. To this end, integration by parts
 yields that
\begin{eqnarray}\label{Z integration by parts new}
\int_{(0,\,u\,]}\varphi(t)\,dZ^\alpha_t=\varphi(t)Z^\alpha_t-\varphi(0)Z_0^\alpha-\int_0^uZ^\alpha_t\varphi'(t)\,dt.
\end{eqnarray}
Since $\alpha$ is bounded, so is $Z^\alpha$. Moreover, $\varphi$ and $\varphi'$ are bounded as well. It hence follows that
 $\int_{(0,\,\cdot\,]}\varphi(t)\,dZ^\alpha_t$ is bounded. But all other terms in
\eqref{Xalpha strategy} are bounded by assumption. Therefore $X$ is an admissible strategy. \qed

\bigskip

\begin{lemma}\label{Xi lemma} Suppose that $M$ is a given constant, $\alpha$ is a   semimartingale satisfying $\bE[\,\int_0^T\alpha_t^2\,dt\,]\le M$, and  $Y^\alpha$, $Z^\alpha$, and $X$ are as in Lemma~\ref{CostLemma}.  Then there exists a constant $C$ that depends only on $M$, $x$, $\rho$, and $T$ such that
$$\bE\bigg[\,\sup_{0\le t< T}\Big( X_t-\frac1{2\rho}\alpha_t\Big)^2\,\bigg]\le C.
$$
Moreover, $|Y_0^\alpha|\le  (1+\rho T)\sqrt{MT}$.
\end{lemma}

\Proof We get from \eqref{Lemma 2 inequality} that 
$\bE[\,(Z_T^\alpha)^2\,]\le MT(1+\rho T)^2$.
Doob's quadratic maximal inequality therefore yields that $Z^*:=\sup_{0\le t\le T}|Z^\alpha_t|$ satisfies
$\bE[\,(Z^*)^2\,]\le 4MT(1+\rho T)^2$. We furthermore have
$$ Y_0^\alpha=Z_0^\alpha\le  \sqrt{\bE[\,(Z^\alpha_T)^2\,]}\le (1+\rho T)\sqrt{MT}$$
 and
$$\sup_{0\le t\le T}|Y^\alpha_t|\le Z^*+(1+2\rho T)\sqrt{T\int_0^T\alpha_t^2\,dt},
$$
and so
$$\bE\big[\,\sup_{0\le t\le T}(Y^\alpha_t)^2\,\big]\le 8MT(1+\rho T)^2+2(1+2\rho T)^2MT.
$$
Next, we get from \eqref{Z integration by parts new} that
$$\sup_{0\le t\le T}\Big|\int_{(0,t]}\varphi(s)\,dZ^\alpha_s\Big|\le  2 Z^*.
$$
Whence,
$$\bE\Big[\,\sup_{0\le t\le T}\Big(\int_{(0,t]}\varphi(s)\,dZ^\alpha_s\Big)^2\,\Big]\le 16MT(1+\rho T)^2.
$$
Since \eqref{Xalpha strategy} holds for $0\le t<T$, we now easily get the assertion. \qed

\bigskip

We will say that $\alpha$ is  a \emph{bounded elementary  process} if it is  of the form
$$\alpha_t=\alpha_0\eins_{\{0\}}(t)+\sum_{i=1}^N\alpha_i\eins_{[\tau_i,\tau_{i+1})}(t),$$
 where $N\in\bN$, the $(\tau_i)$ are stopping times with $0\le\tau_1\le\cdots\le\tau_{N+1}<\infty$, and the coefficients $\alpha_i$ are bounded $\cF_{\tau_i}$-measurable random variables.

\bigskip

\begin{lemma}\label{Cost approx lemma}Let $\alpha$ be a bounded elementary process, $x\in\bR$, $T>0$, and consider the corresponding strategy $X\in\cXsem(x,T)$ constructed in Lemma~\ref{alpha strategy lemma}. Then for each $\eps>0$ there exists a  strategy $\wt X\in\cXBV(x,T)$ such that $|\bE[\,\cC(X)\,]-\bE[\,\cC(\wt X)\,]|<\eps$.
\end{lemma}

\Proof  First, we recall from \eqref{Z integration by parts new} that $N_u:=\int_{(0,u]}\varphi(t)\,dZ^\alpha_t$ is a bounded c\`adl\`ag martingale with $N_0=0$. We set $N_t=0$ for $t<0$ and define
$$N^n_t:=n\int_{t-\frac1n}^tN_s\,ds, \qquad n\in\bN
$$
Then $N^n_t$ is continuous, bounded uniformly in $n$ and $t$, and of bounded variation in $t$. Furthermore, $N^n_t\to N_{t-}$  for all $t\ge0$  as $n\ua\infty$.
Thus, when defining $X^n_{0-}:=x$ and
 $X^n_t:=X_{t}+\frac12\big(N_t-N^n_t\big)$  we have $X^n_{t-}\to X_{t-}$ for all $t$ boundedly. Moreover, we get from \eqref{Xalpha strategy} that, for $0\le t<T$,
\begin{align*}
X^n_t&=\frac{x(1+\rho( T-t)) -\frac{1}2 (1+\rho t)Y_0^\alpha}{2+\rho T}-\frac12N^n_t+\frac1{2\rho}\alpha_t-\rho\int_0^t\bigg(N_s+\frac12\int_0^s\alpha_r\,dr\bigg)\,ds,
\end{align*}
and so $X^n$ is of bounded variation.

Now we set $E^n_t:=\int_{[0,t]}e^{-\rho(t-s)}\,dX_s^n$. Integrating by parts as in \eqref{EtSDE} yields
$E^n_t = X^n_t -x- \rho \int_0^t E^n_s \,ds$.
Therefore,
$$|E^n_{t-}-E_{t-}|\le|X^n_{t-}-X_{t-}|+\rho\int_0^t|E^n_{s-}-E_{s-}|\,ds,
$$
and so Gronwall's inequality (in the extended form of, e.g., Lemma 2.7 in \cite{Teschl}) implies that
$$|E^n_{t-}-E_{t-}|\le |X^n_{t-}-X_{t-}|+ \rho \int_0^t e^{\rho s}     |X^n_{s-}-X_{s-}|\,ds.
$$
Thus, also $E^n_{t-}\to E_{t-}$ boundedly.

With $|A|_{[0,t]}$ denoting the  total variation of $A$ over $[0,t]$, we get from \eqref{ExpectedCostsAlphaEq} that
\begin{eqnarray}\lefteqn{\big|\bE[\,\cC(X^n)\,]-\bE[\,\cC(X)\,]\big|}\nonumber\\
&\le&\bE\Big[\,\int_0^T|X^n_t-X_t||\alpha_t|\,dt+\int_{[0,T]}|X^n_{t-}-X_{t-}|\,d|A|_{[0,t]}\,\Big]\nonumber\\
&&\quad +\rho\Bigg|\bE\bigg[\int_0^{T}\bigg\{\varphi(t)X^n_t+(1-\varphi(t))E^n_t+\frac12\varphi(t)Y^\alpha_t-\frac1{2\rho}\alpha_t\bigg\}^2\,dt\,\bigg]\label{Cost difference aux eq}\\
&&\qquad-\bE\bigg[\int_0^{T}\bigg\{\varphi(t)X_t+(1-\varphi(t))E_t+\frac12\varphi(t)Y^\alpha_t-\frac1{2\rho}\alpha_t\bigg\}^2\,dt\,\bigg]\Bigg|.\nonumber
\end{eqnarray}
Dominated convergence implies that the right-hand side converges to zero when $n\ua\infty$. \qed

\begin{lemma}\label{infinite cost when not abs Lemma}Fix $T>0$ and suppose that $A$ is not $\bP$-a.s. absolutely continuous on $[0,T)$. That is, $A$ is not $\bP$-a.s.  of the form $A_t=\int_0^tA_s'\,ds$ for some progressively measurable  process $A'$ and $0\le t<T$. Then, for any $x\in\bR$,
$$\inf_{\cXBV(x,T)}\bE[\,\cC(X)\,]=-\infty.
$$
\end{lemma}

\Proof Let us define two finite measures $Q $ and $Q^A$ on $([0,T)\times\Om,\mathcal{B}[0,T)\otimes\cF)$ by
$$\int f\,dQ=\int\int_0^Tf(t,\omega)\,dt\,\bP(d\omega),\qquad \int f\,dQ^A=\int\int_{[0,T)}f(t,\omega)\,dA_t(\omega)\,\bP(d\omega),
$$
where $f$ is a bounded measurable function on $([0,T)\times\Om,\mathcal{B}[0,T)\otimes\cF)$. Since $A$ is not absolutely continuous, there exists a bounded measurable function $\bbar\psi\ge0$ on $[0,T)\times\Omega$ such that $\int \bbar\psi\,dQ=0$ and $\int\bbar\psi \,dQ^A=1$. By the predictability of $A$ and Theorem 57 in Chapter VI of \cite{DellacherieMeyerB}, we may replace $\bbar\psi$ by its predictable projection, $\psi$, and still have $\int \psi\,dQ=0$ and $\int\psi\, dQ^A=1$.

It follows from Theorem II.4.10 in \cite{Protter} and a monotone class argument that the left-hand limits of bounded elementary processes are dense with respect to $(Q+Q^A)$-a.e. convergence in the class of predictable processes. Moreover, bounded elementary processes are clearly of finite total variation. By approximating $(K+1) \psi$ for some $K\in\bN$, we hence get that  there exists a bounded elementary process $\alpha\ge0$ such that
\begin{equation}\label{alpha integral le 1}
1\ge \int\alpha^2_-\,dQ=\bE\Big[\,\int_0^T\alpha_{t-}^2\,dt\,\Big]=\bE\Big[\,\int_0^T\alpha_{t}^2\,dt\,\Big]
\end{equation}
 and
$$K\le \int\alpha_-\,dQ^A=\bE\Big[\,\int_{[0,T]}\alpha_{t-}\,dA_t\,\Big].$$

Now let $X\in\cXsem(x,T)$ be the corresponding strategy constructed in Lemma~\ref{alpha strategy lemma}. We denote by $\Xi$ the random variable
\begin{equation}\label{Xi def Eq}
\Xi:=\sup_{0\le t< T}\Big| X_t-\frac1{2\rho}\alpha_t\Big|=\sup_{0<t\le T}\Big|X_{t-}-\frac1{2\rho}\alpha_{t-}\Big|.
\end{equation}
By Lemma~\ref{Xi lemma}, $\bE[\,\Xi^2\,]$ is bounded by a constant $C$ that depends only on $x$, $\rho$, and $T$.
By Lemma~\ref{CostLemma}, \eqref{XoptimalitycondAlphaEq}, and \eqref{Xi lemma}, the expected costs of the strategy $X$ can be estimated as follows:
\begin{eqnarray}\bE \left[\, \cC(X)\, \right]&=&-xS_0+\varphi(0){x^2} + \varphi(0){xY^\alpha_0} - \rho\bE \left[\,  \int_0^T \left(\frac12 \varphi(s){Y^\alpha_s}- \frac1{2\rho} \alpha_s\right)^2 \,ds \,\right]\nonumber\\
&&\qquad+\bE\Big[\,\int_0^TX_t\alpha_t\,dt-\int_{(0,T]}X_{t-}\,dA_t\,\Big]\nonumber\\
&\le&-xS_0+x^2+x(1+\rho T)\sqrt{MT}+\bE\Big[\,\frac1{2\rho}\int_0^T\alpha_t^2\,dt+\Xi\int_0^T|\alpha_t|\,dt\,\Big]\label{Xi cost estimate}\\
&&\qquad- \bE\Big[\,\frac1{2\rho}\int_{[0,T)}\alpha_{t-}\,dA_t-\Xi |A|_{[0,T]}\,\Big]\nonumber\\
&\le&\wt C-\frac{K}{2\rho},\nonumber
\end{eqnarray}
where $|A|_{[0,T]}$ denotes again the total variation of $A$ on $[0,T]$ and $\wt C$ is a constant depending only on $x$, $\rho$,  $T$, and $\bE[\,|A|_{[0,T]}^2\,]$. Here we have also used \eqref{alpha integral le 1}. Since $K$ was arbitrary, it follows that $\inf\bE[\,\cC(X)\,]=-\infty$ when the infimum is taken over $X\in\cXsem(x,T)$. An application of Lemma~\ref{Cost approx lemma} shows that $\cXsem(x,T)$ can be replaced by $\cXBV(x,T)$.\qed

\bigskip

\begin{lemma}\label{infinite cost when not integrable Lemma}Fix $T>0$ and suppose that  $A$ is  $\bP$-a.s.  of the form $A_t=\int_0^tA_s'\,ds$ for some progressively measurable process $A'$ but that $\bE[\,\int_0^T(A'_t)^2\,dt\,]=\infty$. Then, for any $x\in\bR$,
$$\inf_{\cXBV(x,T)}\bE[\,\cC(X)\,]=-\infty.
$$
\end{lemma}

\Proof We have
$$\infty=\sqrt{\bE\Big[\,\int_0^T(A'_t)^2\,dt\,\Big]}=\sup_\psi\bE\Big[\,\int_0^T\psi_tA'_t\,dt\,\Big],$$
where the supremum is taken over all progressively measurable $\psi$ with $\bE[\,\int_0^T\psi^2_t\,dt\,]\le 1$.  By a monotone class argument, the supremum over these $\psi$ can be replaced by a supremum over all bounded elementary processes $\alpha$ with $\bE[\,\int_0^T\alpha^2_t\,dt\,]\le 1$. For every $K>0$ there hence exists a bounded elementary process $\alpha$ such that
\begin{equation}\label{second alpha bounds}
\bE\Big[\,\int_0^T\alpha^2_t\,dt\,\Big]\le 1\qquad\text{and}\qquad \bE\Big[\,\int_0^T\alpha_tA'_t\,dt\,\Big]\ge K
\end{equation}
Let $X\in\cXsem(x,T)$ be the corresponding strategy constructed in Lemma~\ref{alpha strategy lemma} and define $\Xi$ as in \eqref{Xi def Eq}.   Then, due to \eqref{second alpha bounds},
\begin{eqnarray}\bE \left[\, \cC(X)\, \right]&=&-xS_0+\varphi(0){x^2} + \varphi(0){xY^\alpha_0} - \rho\bE \left[\,  \int_0^T \left(\frac12 \varphi(s){Y^\alpha_s}- \frac1{2\rho} \alpha_s\right)^2 \,ds \,\right]\nonumber\\
&&\quad+\bE\Big[\,\int_0^TX_t(\alpha_t-A'_t)\,dt\,\Big]\nonumber\\
&\le&-xS_0+x^2+x\sqrt{MT(1+\rho T)^2}+\bE\Big[\,\frac1{2\rho}\int_0^T\alpha_t^2\,dt+\Xi\int_0^T|\alpha_t|\,dt\,\Big]\nonumber\\
&&\qquad- \bE\Big[\,\frac1{2\rho}\int_{[0,T]}\alpha_{t}A'_t\,dt-\Xi |A|_{[0,T]}\,\Big]\le\wt C-\frac{K}{2\rho},\nonumber
\end{eqnarray}
where $\wt C$ is a constant depending only on $x$, $\rho$,  $T$, and $\bE[\,|A|_{[0,T]}^2\,]$. Since $K$ was arbitrary, it follows that $\inf\bE[\cC(X)\,]=-\infty$ when the infimum is taken over $X\in\cXsem(x,T)$. An application of Lemma~\ref{Cost approx lemma} shows that $\cXsem(x,T)$ can be replaced by $\cXBV(x,T)$.\qed

\bigskip\noindent{\bf Proof of Theorem~\ref{GenSemThm}.} By Lemmas~\ref{infinite cost when not abs Lemma} and~\ref{infinite cost when not integrable Lemma} we may concentrate on the case in which $A$ is absolutely continuous on $[0,T)$ with square-integrable derivative $A'$. Taking $\alpha_t:=A'_t$ in Lemma~\ref{CostLemma} yields that for any strategy $X$,
\begin{equation}\label{ExpectedCostsA'Eq}\begin{split}
\bE \left[\, \cC(X)\, \right]&=-xS_0+\varphi(0){x^2} + \varphi(0){xY_0} - \rho\bE \left[\,  \int_0^T \left(\frac12 \varphi(s){Y_s}- \frac1{2\rho} A'_s\right)^2 \,ds \,\right]\\
&\quad +\rho\bE\bigg[\int_0^{T}\bigg\{\varphi(t)X_t+(1-\varphi(t))E_t+\frac12\varphi(t)Y_t-\frac1{2\rho}A'_t\bigg\}^2\,dt\,\bigg]\\
&\ge -xS_0+\varphi(0){x^2} + \varphi(0){xY_0} - \rho\bE \left[\,  \int_0^T \left(\frac12 \varphi(s){Y_s}- \frac1{2\rho} A'_s\right)^2 \,ds \,\right]
\end{split}
\end{equation}
with equality if and only if \eqref{XoptimalitycondAlphaEq} holds with $\alpha=A'$. Since $X$ was arbitrary, we have
\begin{align*}
\inf_{X\in \cXsem(x,T)} \bE[\cC(X)] \ge \varphi(0){x^2} + \varphi(0){xY_0} - \rho\bE \left[\,  \int_0^T \left(\frac12 \varphi(s){Y_s}- \frac1{2\rho} A'_s\right)^2 \,ds \,\right].
\end{align*}

To show the converse inequality, take bounded elementary processes $\alpha^n$ such that
$$\delta_n:=\bE\Big[\,\int_0^T(\alpha^n_t-A'_t)^2\,dt\,\Big]\longrightarrow 0.$$
 In particular there exists a constant $M$ such that for all $n$ we have $\bE[\,\int_0^T(\alpha^n_t)^2\,dt\,]\le M$. Let $X^n$ be the strategy constructed for $\alpha^n$ in Lemma~\ref{alpha strategy lemma}.  By Lemma~\ref{CostLemma} we have
\begin{equation}\label{C(Xn) approximation eq}\begin{split}\bE \left[\, \cC(X^n)\, \right]&=-xS_0+\varphi(0){x^2} + \varphi(0){xY^{\alpha^n}_0} - \rho\bE \left[\,  \int_0^T \left(\frac12 \varphi(s){Y^{\alpha^n}_s}- \frac1{2\rho} \alpha^n_s\right)^2 \,ds \,\right]\\
&\quad+\bE\Big[\,\int_0^TX^n_t(\alpha^n_t-A'_t)\,dt\,\Big].\end{split}
\end{equation}

By Jensen's inequality, we have
\begin{align*}
\bE\bigg[\,\bigg(\sup_{t\le T}\Big|\int_0^t\alpha_s^n\,ds-\int_0^tA'_s\,ds\Big|\bigg)^2\,\bigg]&\le T\delta_n,\\
\bE\bigg[\,\bigg(\sup_{t\le T}\Big|\int_0^t\int_0^s\alpha_r^n\,dr\,ds-\int_0^t\int_0^sA'_r\,dr\,ds\Big|\bigg)^2\,\bigg]&\le T^3\delta_n.
\end{align*}
In particular, we have $\bE[\,(Z^{\alpha^n}_T-Z_T)^2\,]\to0$.
It therefore follows from Doob's quadratic maximal inequality that $\sup_{t\le T}|Z_t^{\alpha^n}-Z_t|\to0$ in $L^2(\bP)$ and hence moreover that
 $\sup_{t\le T}|Y_t^{\alpha^n}-Y_t|\to0$ in $L^2(\bP)$.
Consequently, the right-hand side of the first line in \eqref{C(Xn) approximation eq} converges to
$$-xS_0+\varphi(0){x^2} + \varphi(0){xY_0}- \rho\bE \left[\,  \int_0^T \left(\frac12 \varphi(s){Y_s}- \frac1{2\rho} A'_s\right)^2 \,ds \,\right].
$$
Furthermore, by the Cauchy--Schwarz inequality,
\begin{eqnarray*}\Big|\bE\Big[\,\int_0^TX^n_t(\alpha^n_t-A'_t)\,dt\,\Big]\Big|\le\sqrt{\delta_n\cdot\bE\Big[\,\int_0^T(X^n_t)^2\,dt\,\Big]}.
\end{eqnarray*}
It is a consequence of Lemma~\ref{Xi lemma} that $\bE[\,\int_0^T(X^n_t)^2\,dt\,]$ is bounded uniformly in $n$, and so it follows that $\bE[\,\int_0^TX^n_t(\alpha^n_t-A'_t)\,dt\,]\to0$. We thus have proved that
$$\bE \left[\, \cC(X^n)\, \right]\longrightarrow -xS_0+\varphi(0){x^2} + \varphi(0){xY_0} - \rho\bE \left[\,  \int_0^T \left(\frac12 \varphi(s){Y_s}- \frac1{2\rho} A'_s\right)^2 \,ds \,\right].
$$
This completes the proof of our formula for the optimal expected costs.

As for optimal strategies, we have already remarked above that equality in \eqref{ExpectedCostsA'Eq} can hold only when
\begin{equation}\label{optimality condition X semimart}
0=\varphi(t)X_t+(1-\varphi(t))E_t+\frac12\varphi(t)Y_t-\frac1{2\rho}A'_t\qquad\text{$\bP$-a.s. for all $t\in[0,T)$.}
\end{equation}
By Lemma~\ref{alpha strategy lemma}, there exists a unique strategy in $\cXsem(x,T)$ satisfying this condition when $A'$ is a bounded semimartingale. This strategy is given by \eqref{X optimal strategy}. When $A'$ is not a semimartingale, it is clear that  \eqref{optimality condition X semimart} cannot be satisfied by a semimartingale $X$. Also, $X$ will be of finite variation if and only if $\varphi(t)Y_t-\frac1{2\rho}A'_t$ is of finite variation. This concludes the proof of Theorem~\ref{GenSemThm}.\qed

\bigskip

\noindent{\bf Proof of Corollary~\ref{A0 martingale corollary}:}
A computation shows that
\begin{align*}
    - Z_t = (1+\rho(T-t))A_t + \frac12(2+\rho(T-t))(T-t)A'_t + \rho\int_0^tA_s\,ds,
\end{align*}
and so in particular
$Y_0= Z_0  = -\frac12(2+\rho T)TA'_0$. From the integration-by-parts formula, $dZ_t$ must satisfy
$$ - dZ_t = \frac12(2+\rho(T-t))(T-t)\,dA'_t+\Lambda_t\,dt,
$$
for a suitable process $\Lambda_t$. But both $Z_t$ and $A'_t$ are martingales, and so we must have $\Lambda_t=0$ (alternatively, the fact $\Lambda_t=0$ can also be verified by a direct computation). Consequently,
$$-\int_{(0,s]}\varphi(u)\,dZ_u=\frac12 \int_{(0,s]}(T-u)\,dA'_u = \frac12\Big((T-s)A'_s - TA'_0 + A_s\Big).
$$
This identity yields that
\begin{align*}
    -\int_0^t\int_{(0,s]}\varphi(u)\,dZ_u\,ds  = \frac12\bigg((T-t)A_t  - TA'_0t + 2\int_0^tA_s\,ds\bigg).
\end{align*}
Plugging these formulas into \eqref{X optimal strategy} yields the assertion after a short computation. \qed

\bigskip

\noindent{\bf Proof of Corollary~\ref{cost-risk corollary}.} We start by simplifying the cost-risk functional \eqref{GS risk functional}. First,  we can assume $S_{0-}^0=0$ without loss of generality and therefore write
$$ \int_{[0,T]} S^0_{t-} \,dX_t + [S^0,X]_T=-\int_{[0,T]}X_{t-}\,dS^0_t
$$
as in \eqref{wt CX integration by parts}.
Then we can write $S^X_t=S^0_t+E_{t-}$. When defining
$$\wh S_t:=S^0_t-\lambda\int_0^tS^0_s\,ds,
$$
we can thus write
\begin{equation}\label{cost-risk simplified}\begin{split}
\lefteqn{\bE\Big[\,\cC(X)+\lambda\int_0^TS^X_tX_t\,dt\,\Big]}\\
&= \bE\left[\, -\int_{[0,T]} X_{t-} \,d\wh S_t  + \int_{[0,T]} E_{t-} \,dX_t + \frac12 [X]_T +\lambda\int_0^TE_tX_t\,dt\,\right].
\end{split}\end{equation}
To simplify \eqref{cost-risk simplified} further, we
let
$$\bbar E_t:=\int_0^t E_s\,ds=\frac1\rho\Big(X_t-x-E_t\Big),$$
where we have used \eqref{EtSDE} in the second step, and set $\bbar E_{0-}=0$. Then \begin{eqnarray*}\int_0^TE_tX_t\,dt&=&\int_{[0,T]}X_t\,d\bbar E_t=X_T\bbar E_T-X_{0-}\bbar E_{0-}-\int_{[0,T]}\bbar E_{t-}\,dX_t\\
&=&\frac1\rho\int_{[0,T]} E_{t-}\,dX_t-\frac1\rho\int_{[0,T]} X_{t-}\,dX_t-\frac{x^2}{\rho}\\
&=&\frac1\rho\bigg(\int_{[0,T]} E_{t-}\,dX_t+\frac12[X]_T-\frac12x^2\bigg).
\end{eqnarray*}
It follows that
\begin{align*}
\lefteqn{\bE\Big[\,\cC(X)+\lambda\int_0^TS^X_tX_t\,dt\,\Big]}\\
&=  -\frac{\lambda}{2\rho}x^2+\Big(1+\frac\lambda\rho\Big)\bE\left[\,\int_{[0,T]} \wt S^0_{t-} \,dX_t + [\wt S^0,X]_T + \int_{[0,T]} E_{t-} \,dX_t + \frac12 [X]_T \,\right],
\end{align*}
where
$$\wt S^0_t=\frac{\wh S_t}{1+\frac\lambda\rho}=\frac\rho{\rho+\lambda}\bigg(S^0_t-\lambda\int_0^tS^0_s\,ds\bigg).
$$
This concludes the proof. \qed

\medskip

\noindent{\bf Acknowledgement:} We wish to thank Markus Hess and two anonymous referees for helpful comments on a previous version of the manuscript.

\parskip-0.5em\renewcommand{\baselinestretch}{0.9}\small
\bibliography{MarketImpact}{}
\bibliographystyle{agsm}

\end{document}